\documentclass[aps,prl,groupedaddress,twocolumn
]{revtex4-1}  
\usepackage{graphicx,bm,amssymb}  

\def\be{\begin{equation}}  
\def\ee{\end{equation}}  
\def\ba{\begin{eqnarray}}  
\def\ea{\end{eqnarray}}  
\def\bc{\begin{center}}  
\def\ec{\end{center}}

\begin{document}

\title{Ferroelectric instability of two-dimensional crystals
}

\author{S. A. Mikhailov }

\email[Electronic mail: ]{sergey.mikhailov@physik.uni-augsburg.de}  

\affiliation{Institute of Physics, University of Augsburg, D-86135 Augsburg, Germany}  

\date{\today}  

\begin{abstract}
The macroscopic dielectric permittivity of dielectric crystals is related to the microscopic atomic polarizability of constituent atoms by the known Clausius-Mossotti relation obtained in the middle of 19th century. We derive a similar relation for recently discovered \textit{two-dimensional} crystals (mono- and bilayer graphene, boron nitride, etc) and show that, in contrast to three-dimensional materials, much stronger electron-electron interaction in two dimensions leads to a spontaneous electric polarization of the ground state of two-dimensional crystals. The predicted ferroelectric transition may have interesting applications in electrodynamics and optics.
\end{abstract}

\maketitle  

The macroscopic dielectric susceptibility $\chi$ and permittivity $\epsilon$ of dielectric crystals are related to the microscopic atomic polarizability $\alpha$ of constituent atoms by the known Clausius-Mossotti relation \cite{Clausius1879,Mossotti1850},
\be 
\frac{\epsilon-1}{4\pi}=\chi=\frac{N_v\alpha}{1-4\pi N_v\alpha/3},\label{ClausMos}
\ee
where $N_v$ is the volume concentration of atoms. The divergence of $\chi$ and $\epsilon$ at $N_v\alpha\to 3/4\pi$ is known as the polarization catastrophe (e.g., \cite{Bhatt84,Kittel05}), which leads to a ferroelectric instability of the ground state of the crystal. Physically, this is a consequence of the local field effects \cite{Kittel05}: the electric field, which acts on each atom of the crystal and polarizes it, differs from the external one by the fields produced by all other polarized atoms in the crystal lattice. 

\begin{figure}
\includegraphics[width=\columnwidth]{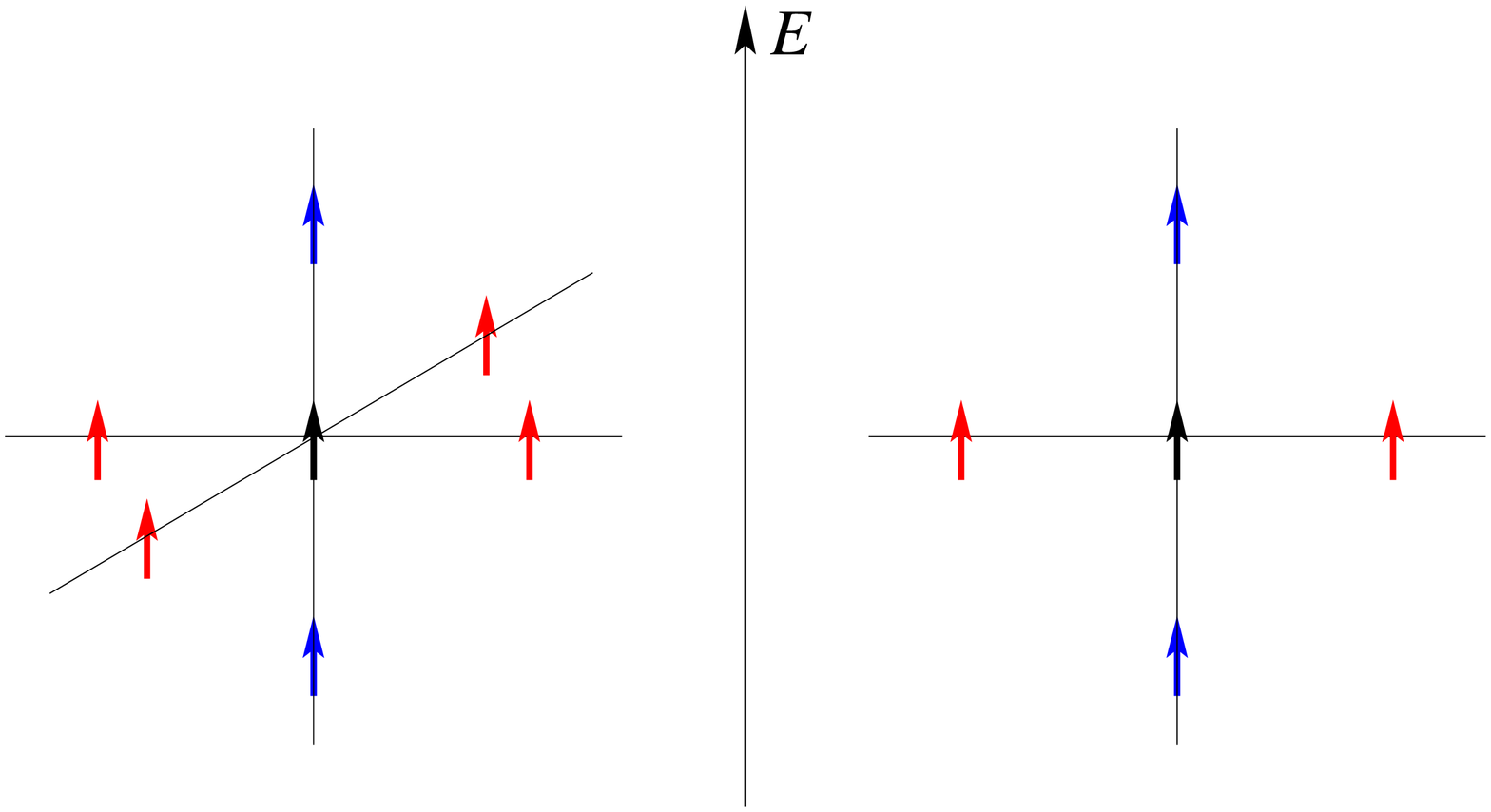}
\caption{\label{cublatt}The local field effects in (left) three- and (right) two-dimensional simple cubic/square lattices. The electric field $-4d/a^3$ of the four red dipoles in the 3D cubic lattice (left) exactly compensates the field $+4d/a^3$ of the two blue dipoles. In the 2D square lattice (right) there are only two red dipoles and the resulting local field is $+2d/a^3$. Its direction coincides with the direction of the external field. }
\end{figure}

In three-dimensional (3D) crystals, however, the influence of the local fields is not very strong since they often cancel each other. This is illustrated in Fig. \ref{cublatt}(left) for the case of a simple cubic lattice. In such a lattice each dipole is surrounded by four nearest neighbors in the azimuthal plane and two neighbors in the vertical direction. Since the dipole field 
\be 
{\bm E}_{dip} ({\bm  r})
=\frac{3({\bm d}\cdot{\bm r}){\bm  r}-{\bm  d} r^2}{r^5}
\ee
is strongly anisotropic, four azimuthal (red) dipoles create the field $-4\times \bm d/a^3$ opposite to the external one, while the two blue (``north'' and ``south'') dipoles produce the field $+2\times 2\bm d/a^3$ in the same direction as the external field ($a$ is the lattice constant). The sum of these fields vanishes. 

The discovery of graphene \cite{Novoselov04,Novoselov05,Zhang05} and other atomically thin crystals \cite{Novoselov05Nat} opened a way of exploiting new types of materials -- {\em two-dimensional} crystals. As seen from Fig. \ref{cublatt}, in purely two-dimensional crystals the local field is much stronger, since two azimuthal (red) dipoles are absent. This local field is really huge; for example, if electrons are shifted from their host atoms by only $\delta x\simeq 0.01$ \AA, the field $2d/a^3=2e\delta x/a^3$ from the nearest four dipoles shown in Fig \ref{cublatt}(right) is about $2\times 10^6$ V/cm (for a typical lattice constant $a\simeq 2.5$ \AA). 

\begin{figure}
\includegraphics[width=0.8\columnwidth]{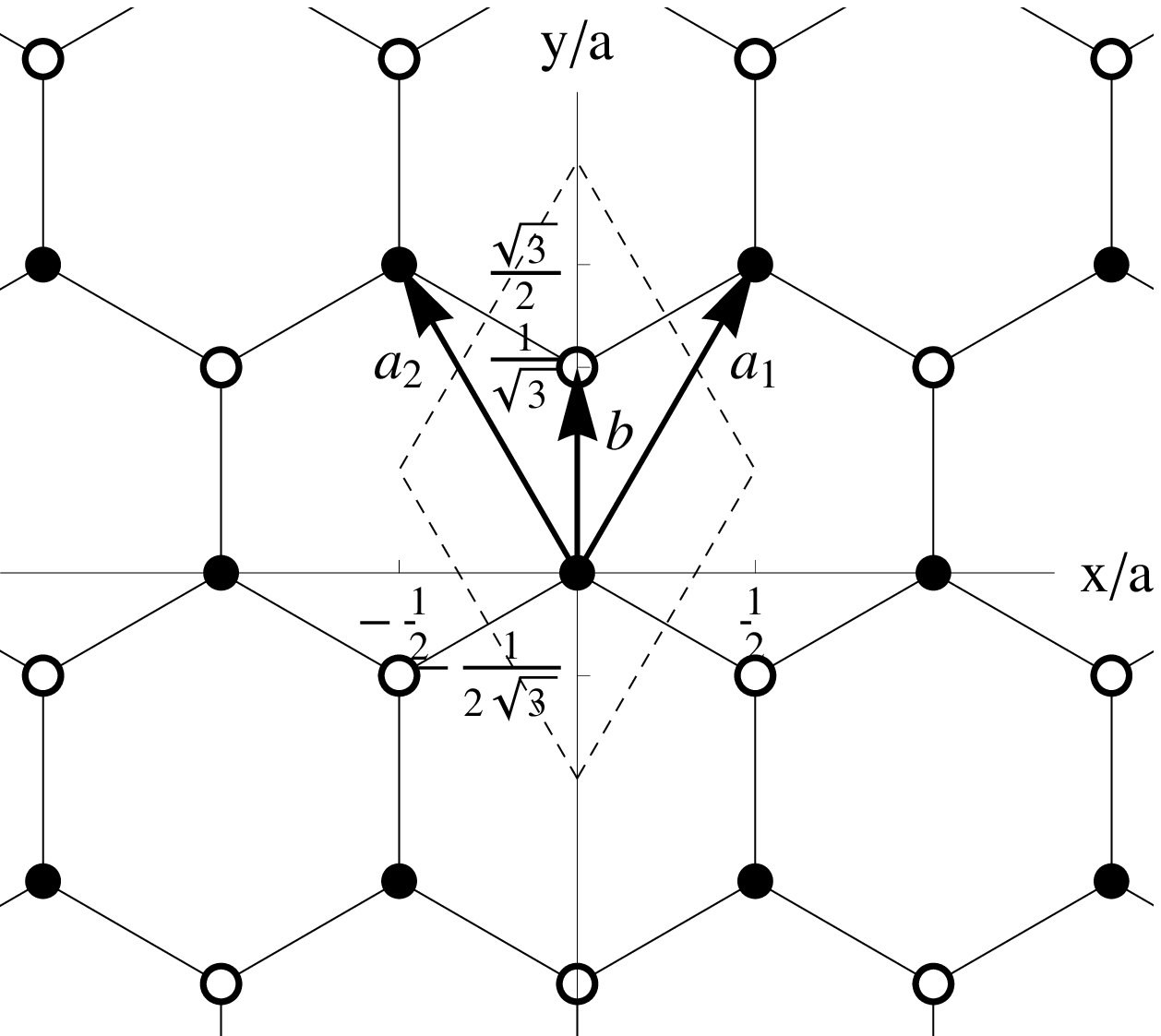}
\caption{\label{lattice}The hexagonal two-dimensional lattice of graphene and boron nitride. }
\end{figure}

The real 2D crystals (graphene, boron nitride) have a hexagonal lattice, Fig. \ref{lattice}, consisting of two triangular sublattices $A$ and $B$ (black and open circles). If the external field $\bm E_0$ is parallel to the 2D plane, the induced dipole moments $d_A$ and $d_B$ satisfy the equations
\be 
d_A
=\alpha_A\left\{ E_0
+\frac{d_A}{2a^3} S_{AA}
+\frac{d_B}{2a^3}S_{AB}
\right\},
\ee
\be
 d_B
=\alpha_B\left\{ E_0
+\frac{d_A}{2a^3}S_{BA}
+\frac{d_B}{2a^3}S_{BB}
\right\},
\ee
where $\alpha_A$ and $\alpha_B$ are atomic polarizabilities of the $A$ and $B$ atoms, and $a$ is the lattice constant. The sums
\be 
S_{AA}=S_{BB}=
\sum_{(m,n)\neq (0,0)}\frac{1}{(m^2+mn +n^2)^{3/2}}\approx 11.034,\label{sumAA}
\ee
and
\ba
S_{AB}=S_{BA}&=&
\sum_{m,n}\frac{1}{(m^2+mn +n^2+m+n+\frac 13)^{3/2}}
\nonumber \\ &\approx& 23.151.\label{sumAB}
\ea 
correspond to the summation over all dipoles of the same and of the other sublattice. The susceptibility of the hexagonal 2D lattice then assumes the form
\be 
\chi=\frac{N_s}2\frac{\alpha_A+\alpha_B +(\alpha_A\beta_B+\alpha_B\beta_A)  (S_{AB}-S_{AA})}
{(1-\beta_A  S_{AA})(1-\beta_B  S_{AA})-\beta_A\beta_B  S_{AB}^2}, 
\label{AneqB}
\ee
where $\beta=\alpha/2a^3$ and $N_{s}=4/\sqrt{3}a^2$ is the surface density of atoms. If the atoms $A$ and $B$ are identical, then $\chi=N_{s}\alpha /[1-\beta  (S_{AA}+S_{AB})]$.

\begin{figure}
\includegraphics[width=0.8\columnwidth]{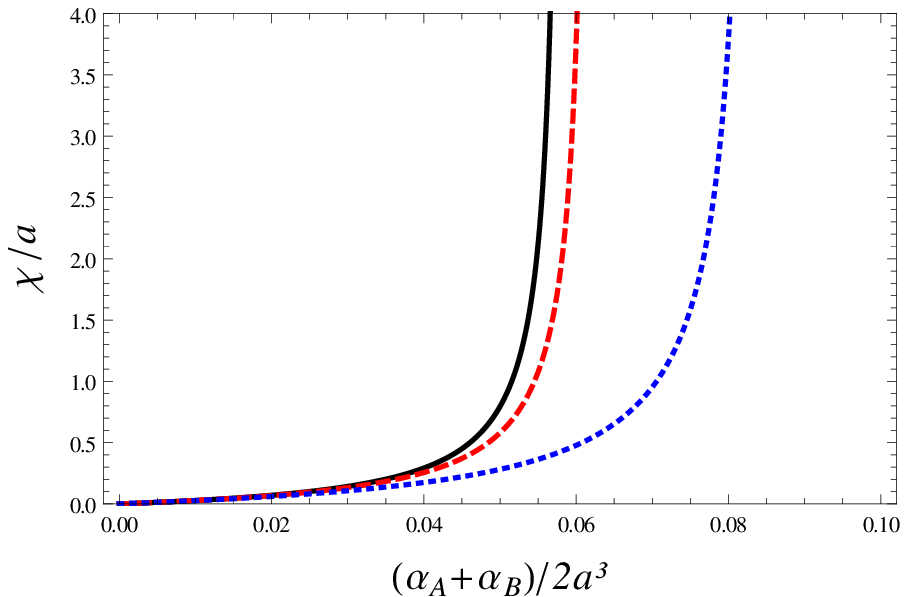}
\caption{\label{instab}The susceptibility of the hexagonal 2D lattice $\chi/a$ as a function of $\bar\alpha/a^3=(\alpha_A+\alpha_B)/2a^3$, at $\eta=1$ (black, solid curve), $\eta=0.78$ (red, dashed) and $\eta=0.1$ (blue, dotted). The black and red curves correspond to parameters of graphene and boron nitride, respectively.}
\end{figure}

Figure \ref{instab} shows the dimensionless susceptibility $\chi/a$, Eq. (\ref{AneqB}), of the 2D hexagonal lattice as a function of $\bar\alpha/a^3=(\alpha_A+\alpha_B)/2a^3$ at several values of the parameter $\eta=4\alpha_A\alpha_B/(\alpha_A+\alpha_B)^2$. When the ratio $\bar\alpha/a^3$ becomes sufficiently large the susceptibility diverges and the system becomes unstable with respect to the spontaneous dielectric in-plane polarization. If the atoms $A$ and $B$ are identical ($\eta=1$, black solid curve in Fig. \ref{instab}), this happens at   
\be 
\frac \alpha {a^3}=\frac 2{S_{AA}+S_{AB}}\approx 0.0585.\label{stab_condition}
\ee
If $\eta=0.78$ (red dashed curve), the system is stable at $\bar\alpha/a^3<0.06225$. One can show that in the monoatomic square lattice the stability boundary lies at $\alpha/ a^3 = 0.2214$, and in a linear chain of atoms, i.e. in a truly one-dimensional crystal, a spontaneous electric polarization along the chain would arise at $\alpha /a^3 \ge 0.2080$.

The predicted ferroelectric transition in the ground state of 2D crystals is a consequence of strong electron-electron ($e$-$e$) interaction. Notice that the tight-binding approximation (TBA), as well as other approaches, which ignore $e$-$e$ interaction or take it into account perturbatively, cannot properly describe the predicted effect. In particular, the ferroelectric ground state is degenerate with respect to the in-plane direction of the spontaneous polarization, while the tight-binding ground state is non-degenerate. On the other hand, the applicability of TBA to 3D crystals \cite{Wallace47} is beyond question since the local field effects are much weaker in three dimensions, Fig. \ref{cublatt}. 

Let us apply the general results obtained above to real 2D crystals with the hexagonal lattice, graphene and boron nitride, BN. Using the atomic polarizability of carbon, $\alpha_C\simeq 1.63 - 1.73$ \AA$^3$, \cite{Schwerdtfeger06} and the lattice constant of graphene, $a=2.46$ \AA, we get $\alpha_C/a^3\gtrsim 0.1095$. For boron nitride ($\alpha_B=3.04$ \AA$^3$, $\alpha_N=1.10$ \AA$^3$ \cite{Schwerdtfeger06}, $a=2.52$ \AA), we get $\eta=0.78$ (corresponds to the red curve in Fig. \ref{instab}) and $\bar\alpha/a^3=0.129$. Both values are far beyond the stability boundaries (0.0585 and 0.06225, respectively). The \textit{suspended} graphene and boron nitride should thus be in the ferroelectric ground state with the spontaneous dielectric polarization of the crystal lattice (the same is valid for bilayer graphene, too). Three-dimensional graphite, in contrast, is stable, as follows from the 3D Clausius-Mossotti formula (\ref{ClausMos}). 

The predicted ferroelectric instability is a peculiar property of two-dimensional crystals. An extension of the system in the third dimension returns it back to a stable state (as seen from the above comparison of mono-/bilayer graphene with three-dimensional graphite). In particular, if a 2D crystal lies on a substrate with the dielectric constant $\epsilon$, screening of the local fields by the substrate may suppress the instability (this may be the reason of why the predicted transition has not been experimentally discovered so far). 
For example, for graphene or BN lying on a SiO$_2$ substrate ($\epsilon_{SiO_2}=3.9$), the polarizability $\bar\alpha$ in the above formulas should be replaced by $\bar\alpha/\epsilon_{eff}$, where $\epsilon_{eff}=(\epsilon+1)/2=2.45$, and the instability conditions are no longer satisfied (for graphene $\alpha_C/a^3\epsilon_{eff}=0.0447<0.0585$; for BN $\bar\alpha/a^3\epsilon_{eff}=0.0526<0.06225$). On the other hand, choosing an appropriate substrate one could put the system very close to the transition point, where the 2D susceptibility is very large, see Fig. \ref{instab}. Optical properties of monolayer dielectric crystals with a very large susceptibility $\chi$ are very interesting and deserve a separate extensive study. For example, a mono-atomic layer with a large susceptibility reflects almost 100\% of incident light. 

\acknowledgments

I would like to thank Kostya Novoselov, Sergey Ganichev, Jonathan Eroms, Oleg Pankratov and Michael Fogler for useful comments and discussions. The financial support of this work by the Deutsche Forschungsgemeinschaft is gratefully acknowledged.


\end{document}